\documentclass[preprint]{aastex}
\usepackage{psfig}
\usepackage{natbib}

\newcommand{\nn}{\nonumber \\}
\newcommand{\ls}{\mathrel{\raise1.16pt\hbox{$<$}\kern-7.0pt 
\lower3.06pt\hbox{{$\scriptstyle \sim$}}}}         
\newcommand{\gs}{\mathrel{\raise1.16pt\hbox{$>$}\kern-7.0pt 
\lower3.06pt\hbox{{$\scriptstyle \sim$}}}}         

\long\def\comment#1{}

\def\fun#1#2{\lower3.6pt\vbox{\baselineskip0pt\lineskip.9pt
  \ialign{$\mathsurround=0pt#1\hfil##\hfil$\crcr#2\crcr\sim\crcr}}}

\newcommand{\ba}{\begin{eqnarray}}
\newcommand{\ea}{\end{eqnarray}}
\newcommand{\be}{\begin{equation}}
\newcommand{\ee}{\end{equation}}
\newcommand{\vk}{{\bf{k}}}

\newcommand{\vx}{{\bf{x}}}

\begin{document}
\title{On the trispectrum as a gaussian test for cosmology}
\author{Licia Verde}
\affil{Depart. of Astrophysical sciences, Peyton
Hall, Princeton University, Ivy lane, Princeton, NJ 08544 1001, USA\\
Institute for Astronomy, University of Edinburgh,
Blackford Hill, Edinburgh EH9 3HJ, UK}
\email{lverde@astro.princeton.edu}
\author{Alan F. Heavens}
\affil{Institute for Astronomy, University of Edinburgh, Blackford Hill,
Edinburgh EH9 3HJ, UK}
\email{afh@roe.ac.uk}

\begin{abstract}
In the standard model for structure formation, bound objects originate
from the gravitational collapse of small perturbations arising from
quantum fluctuations with random phases.  In other scenarios, based on
defects, structures are seeded by localized energy density.  In
principle, it is possible to differentiate between these models on the
basis of their statistical properties; only in the former case is the
initial density field an almost-perfect random gaussian field.  In
this paper, we investigate the use of the trispectrum of the galaxy density field, which is the
connected four-point function in Fourier space, as a discriminant
between gaussian and non-gaussian models.  It has the advantage of
having only weak non-linear growth.  We define a related statistic
$\tau$ which, as a test of the gaussian hypothesis, is independent of
cosmology, the power spectrum and biasing, in real space, and which
is, in principle, a measure of the departure from gaussian statistics.
For galaxy redshift surveys, the statistic depends on cosmology and
bias only through the potentially observable parameter $\beta$.  We
compute the expected errors on the estimate of $\tau$, and demonstrate
with numerical simulations that it can be a useful discriminant of
models, with the important proviso that any bias is linear on large
scales. Whether it is the most effective method is uncertain and
depends on the nature of the departure from gaussianity.
\end{abstract}
\section{Introduction}

Models of primordial fluctuations that generated the large-scale
cosmological structures can be divided into two broad classes: gaussian and
non-gaussian.  The simplest versions of inflationary cosmology predict
almost perfectly gaussian initial fluctuations, but such fluctuations
can arise under more general conditions from the central limit theorem.  The
popularity of gaussian models is also due to their mathematical
simplicity and the possibility they present for analytical
calculations.  On the other hand, density fields generated by
topological defects \citep{vilke85,vach86,hillscrammfry89,Tur89,albsteb92} have
non-gaussian initial conditions; the same can be said about fields
generated by some versions of inflation \citep{allengrinsteinwise87,kofpog88,SBB89}.  Convincing evidence against
gaussian initial conditions (GIC) would rule out many scenarios, and/or
point us towards a physical theory for the origin of primordial
fluctuations.  Thus a test of the gaussian nature of the initial
conditions is of great interest for cosmology.

Microwave background anisotropies probe cosmic fluctuations at a time
when their statistical distribution should be close to its primeval
form. To date, due to the limited signal-to-noise ratio of existing
data, no conclusive evidence about the gaussianity of the initial conditions has been reached (e.g., \cite{Hea98,Ferreira98,Hinshaw94,FRS93,LS93,GLMM94,Luo94b,Smoot94,Kogut96,BandayZG99,TegBrom99}).  An alternative is to analyze the present day
density field of the galaxy distribution. This approach is complicated
by the fact that the density field we observe today has already
undergone non-linear gravitational evolution, and the observed
non-gaussian nature of the galaxy distribution on small scales may be
entirely the result of non-linear gravitational clustering combined
with biasing effects. 
This non-linear growth makes any intrinsic non-gaussian signal more
difficult to detect, and leads one to the conclusion that generally
study of the microwave background is likely to be more profitable in
detecting non-gaussian features \citep{VWHK00}.  However, one cannot
exclude the possibility that non-gaussian features arise on physical scales which
are difficult to probe with the microwave background, or non-GIC might produce a gaussian Sachs-Wolfe effect
(e.g., \cite{sherrershaffer95}), so large-scale structure studies may
still have a role to play in the test of the GIC hypothesis.  An alternative method is to study number
densities of rare/high-redshift objects (e.g., \cite{COS98,RGS99,Willick00,MVJ00,VJKM00,VKMB}).

There have been numerous studies to calculate the departures from
gaussianity induced by gravity, and to set up a
test of the gaussian nature of the  initial conditions.  There are essentially three different
approaches: a) using N-body simulations, starting from GIC and
several non-GIC, compare the resulting
clustering properties of the evolved fields (e.g., \citet{MMLM91},
\citet{MLMM91},
\citet{WC92}) b) investigate the topological properties of fields
generated from gaussian and non-gaussian conditions and set up a
comparison (e.g., \cite{MMLLM90,CMPLMM93,VPGHG94,avelino97}) c) measure the moments of the density or the
velocity fields and compare them with those predicted from a gaussian
distribution or different non-gaussian models (e.g.,
\cite{Fry84,frysher94,sherrer92,CM94b,Cs95}, Gaztanaga \& Fossalba (1998a,b), \cite{KimStrauss98,Bernardeau94b}).

In this paper, we take the last approach, but work in Fourier space.
We concentrate on the 4-point function, for the following reasons.
Under rather general conditions, the two-point function carries no
model-independent information about the gaussian nature of the field
\citep{FanBardeen95}.  The bispectrum (3-point function) is more
promising, as it is zero for a gaussian field, but generally non-zero.
However, it grows at second-order in perturbation theory, and this
signal will place limits on the accuracy with which one can identify a
primordial component \citep{VWHK00}.  The trispectrum (4-point
function) has the advantage that it grows linearly, with contributions
from non-linear growth appearing only weakly in perturbation
theory. This will be quantified in Sec. 2.  One might hope, therefore,
that the linear growth extends to relatively small scales.  In
addition, the possibility is open that non-gaussian models have no
bispectrum, or even negative initial bispectrum (e.g., \cite{MMLM91}).

One can also probe the 4-point function in real space
(e.g., \cite{LS93,lokas95,Chodbouch96}), or look at a subset of configurations  of the four-point function,
through power correlations \citep{FKP94,SP96}.
An alternative way to approach mildly non-gaussian fields is based on
the Edgeworth expansion \citep{amendola96}.  

The main difficulty with these approaches appears if the smoothed
galaxy distribution is related to the underlying mass distribution by
a {\em non-linear} transformation.  In this circumstance, it is
probably very difficult to distinguish non-GIC
from a non-linear bias, so one has to assume that, at least on large
scales, the bias is linear and deterministic (cf., \cite{TMJS01}).  In any eventuality, there is a realistic
possibility that current large galaxy surveys such as the
Anglo-Australian 2-degree Field (hereafter 2dF;
\cite{Colless96}) and the Sloan Digital Sky Survey \citep{Yorketal00} will
place some constraints on any initial departures from gaussian
behavior.

This paper is organized as follows: In section 2 we review the
relevant statistical properties of the density field through its
$n$-point functions, and introduce the $\tau$ statistic as a
non-gaussian discriminant.  Section 3 contains the practical
implementation of the method on numerical simulations.  Redshift-space
distortions are considered in Section 4 and conclusions are presented
in Section 5.

\section{Statistics of the density field in the linear and weakly non-linear
regime}  
The statistical properties of the fractional overdensity field
$\delta(\vx)=[\rho(\vx)-\overline{\rho}]/\overline{\rho}$, can be characterized
by the $n$-point correlation functions or, in Fourier space, by the $n$-point
spectra. If the fluctuation field is gaussian, the connected part of the
$n$-point function vanishes for $n \geq 3 $ (e.g., \cite{bertvalencia92}). Thus the two-point function, or
alternatively the power spectrum, completely specifies a gaussian
distribution.

To linear order in perturbation theory an initially gaussian
distribution remains gaussian; in particular the the connected
$n$-point functions and $n$-point spectra of an initially gaussian
distribution are zero as long as linear perturbation theory holds.  On
the other hand, if the initial conditions are non-gaussian, the
$n$-point spectra in the linear regime are the primordial ones, scaled
by the $n^{th}$-power of the linear growth factor. Therefore it would be
possible to detect primordial non-gaussianity with a measurement of
non-zero connected linear $n$-spectra for $n \geq 3$.

We concentrate here on the four-point function in Fourier space (the
trispectrum), because, as we demonstrate below, second-order
contributions vanish for gaussian primordial fluctuations, so we hope
to be able to use linear theory to smaller scales. This is tested
numerically. Moreover this quantity can be made independent of the
cosmological model, the degree of linear bias, and the power spectrum
at least on large scales as we will show below.

The data which we will work with are products of four Fourier coefficients 
(in practice we use the real part of this):
\be
D_{\alpha} = \delta_{\vk_1}\delta_{\vk_2}\delta_{\vk_3}\delta_{\vk_4}.
\ee
The mean value of this is zero (by homogeneity), unless the four
wavevectors form a quadrilateral. $\alpha$ labels the set of 4
wavevectors. 

Note that the trispectrum itself is strictly the Fourier counterpart of the connected part of the four-point correlation function only. For
a zero-mean field with non-zero connected (subscript $c$) four-point function, 
\ba
 \langle
\delta_{\vk_1}\delta_{\vk_2}\delta_{\vk_3}\delta_{\vk_4}\rangle & =& \nn
 \langle \delta_{\vk_1}\delta_{\vk_2}\rangle\langle
\delta_{\vk_3}\delta_{\vk_4}\rangle &+& \mbox{(2 perms.)}
  +\langle \delta_{\vk_1}\delta_{\vk_2}\delta_{\vk_3}\delta_{\vk_4}
\rangle_{c}  
\label{Wick}
\ea
where
\ba
\langle \delta_{\vk_i}\delta_{\vk_j}\rangle &= & (2\pi)^3
P(k_i)\delta^D(\vk_i+\vk_j)\\\nonumber
\langle \delta_{\vk_1}\delta_{\vk_2}\delta_{\vk_3}\delta_{\vk_4}\rangle_{c}&= & (2\pi)^3
T(\vk_i)\delta^D(\vk_1+\vk_2+\vk_3+\vk_4).
\label{Power}
\ea
Here $P(k)$ is the power spectrum, $T$ is the trispectrum,
$\delta^D(\vk)$ is the three-dimensional Dirac delta function, the
angle brackets indicate the ensemble average and $(\vk_i)$ is a shorthand for $(\vk_1,\vk_2,\vk_3,\vk_4)$.

Note that $\langle\delta_{\vk_1}\delta_{\vk_2}\delta_{\vk_3}\delta_{\vk_4}\rangle$
has a specific volume dependence when one works with a discrete rather
than continuous Fourier transform.  The gaussian part involves
products of two Dirac delta functions, whereas the connected part has
only one.  For a volume-limited survey of volume $V$, the discrete
Fourier transform changes these delta functions to
$\delta^D\longrightarrow V/(2\pi)^3$ multiplied by a Kronecker delta,
so any comparison between the size of non-gaussian and gaussian parts
of $D_{\alpha}$ is volume-dependent.  Any statement about the relative
importance of the terms must therefore be made with some caution.  We
have to stress here that this is a general feature of $\langle
\delta_{\vk_1}\ldots \delta_{\vk_n}\rangle$
that are evaluated in a finite volume: the relative importance of the
connected and non-connected parts is volume dependent.

In order to demonstrate the lack of contribution to $\langle
D_{\alpha}\rangle$ from second-order perturbation theory, we expand
the density field to second order as:
\be
\delta(\vx)=\delta^{(1)}(\vx)+\delta^{(2)}(\vx)
\label{2Odelta}
\ee
where $\delta^{(1)}$ is the linearly-evolved overdensity.
$\delta^{(2)}$ is $O(\delta^{(1)2})$ and, if initial conditions are
gaussian, represents departures from gaussian behavior due to
gravitational evolution (e.g., \cite{Goroff86}, 
\cite{BJCP92}).  Strictly speaking the galaxy distribution is not a
mildly non-linear field, but a highly non-linear field filtered on
some smoothing scale. The operations of smoothing and evolution do not
commute, but experiments (e.g., \cite{MVH97}, hereafter MVH97) show
that perturbation theory works well if the smoothed field is not too
non-linear.

To second order in perturbation theory the Fourier counterpart of the
connected four-point correlation function can be obtained by applying to
eq. (2) the substitution 
\be
 \delta_{\vk}=\delta_{\vk}^{(1)}+\epsilon\delta_{\vk}^{(2)}+\epsilon^2
\delta_{\vk}^{(3)}+\ldots
\label{eq.2opt}
\ee
and retaining only terms $\propto \epsilon$. Here $\epsilon$ is simply  a bookkeeping
parameter that can then be set to unity and the expression for $\delta^{(2)}_{\vk}$ can be found e.g., in
\citet{Fry84} and \citet{CLMM95}. Hence 

\ba
\langle \delta_{\vk_1}\delta_{\vk_2}\delta_{\vk_3}\delta_{\vk_4}\rangle & =   &
 \langle\delta^{(1)}_{\vk_1}\delta^{(1)}_{\vk_2}\delta^{(1)}_{\vk_3}\delta^{(1)}_{\vk_4}\rangle+\nn
& & \langle\delta^{(1)}_{\vk_1}\delta^{(1)}_{\vk_2}\delta^{(1)}_{\vk_3}\delta^{(2)}_{\vk_4}\rangle
+ cyc. \mbox{ (4 terms)}.
\label{2Otrisp}
\ea
There are 4 cyclical terms involving the second order
$\delta^{(2)}_{\vk}$ whose expression involves products of two linear
coefficients, so for an initially gaussian field, the second-order
contributions to the trispectrum are products of five coefficients and
hence vanish. The first non-vanishing contribution to the connected
trispectrum due to gravitational instability is $\propto \epsilon^2$
as is the next (third order) term in (\ref{eq.2opt}).  In general, the
trispectrum will depend on all orders of correlations present
initially, but in the linear regime, the only contribution is from the
initial trispectrum ($T_{initial}$).  Specifically, if ${\cal D}(t)$
is the growth factor:
\be
T=T_{initial} \; {\cal D}(t)^4.
\ee
For GIC, the trispectrum is zero,
but $\langle D_{\alpha}\rangle_{GIC}$ has  a disconnected part
\be
\langle D_{\alpha}\rangle_{GIC}=\!(2\pi)^6\!P(k_1\!)P(k_3\!)\delta^D(\vk_1\!+\!\vk_2\!)
\delta^D(\vk_3\!\!+\!\vk_4\!)\!+cyc.
\label{Dalpha}
\ee
If primordial fluctuations are non-gaussian, then $\langle D_{\alpha}
\rangle \ne \langle D_{\alpha}\rangle_{GIC}$. 
For $k$-vector configurations where all $|\vk|$ are
different, $\langle D_{\alpha}\rangle_{GIC}$ vanishes, leaving $\langle
D_{\alpha}\rangle = (2\pi)^3T(\vk_1,\vk_2,\vk_3,\vk_4)\delta^D$.
We can thus  characterize the  
departure from gaussian behavior by
\ba
& &\langle\delta_{\vk_1}\delta_{\vk_2}\delta_{\vk_3}\delta_{\vk_4}\rangle \nn
&=&(2\pi)^6P(k_1)P(k_2)\delta^D(\vk_1+\vk_3)\delta^D(\vk_2+\vk_4)+cyc. \nn
&+&\tau\sqrt{P(k_1)P(k_2)P(k_3)P(k_4)}\delta^D(\vk_1+\vk_2+\vk_3+\vk_4),
\label{eq:trisptau}
\ea
where we have introduced the quantity
\be
\tau \equiv \frac{T(\vk_1,\vk_2,\vk_3,\vk_4)}{\sqrt{P(k_1)P(k_2)P(k_3)P(k_4)}}.
\label{eq:tau}
\ee  
$\tau$ is independent of the volume, the cosmological model, linear bias and
redshift (or alternatively the power spectrum amplitude), but in general
depends on the $\vk_i$ (i.e. the scale and the configuration of the quadrilateral formed by the 4 $k$-vectors).
Here for simplicity we will use equilateral k-vector configurations; with this
choice isotropy demands that $\tau$ depends only on $|\vk|$, for a given shape.
We can then use a likelihood method to
estimate $\tau(k)$, and deduce whether it is consistent with the gaussian
result ($\tau= 0$).

\subsection{Likelihood}

Our treatment here follows that of MVH97. 
For simplicity, we would like to be
able to adopt a {\it gaussian} likelihood (italics will refer to the
form of the likelihood function, not the initial statistics), which
will be the {\it a posteriori} probability distribution for $\tau$ if we
assume a uniform prior:
\ba
{\cal L}(\tau)& = &\frac{1}{(2 \pi )^{\frac{M}{2}} 
({\rm det} C)^{\frac{1}{2}}}\times \nn 
& & \exp{\left[-\frac{1}{2} \sum_{\alpha \beta}(D_{\alpha } - \mu _{\alpha}) 
C_{\alpha \beta}^{-1} ( D_{\beta} - \mu_{\beta }) \right]} .\nn 
\label{like}
\ea

Here $M$ is the number of data, which have means $\mu_\alpha(\tau) 
\equiv \langle D_\alpha \rangle$, and that can be evaluated using equation 9.  The covariance matrix is  
$C_{\alpha \beta} \equiv \langle(D_{\alpha }- \mu_{\alpha })(D_{\beta} - 
\mu _{\beta })\rangle$. If $\tau=0$ the likelihood analysis is effectively a
$\chi ^2$ analysis.

In the general (i.e. non-gaussian) case we cannot justify the applicability of the
{\it gaussian} likelihood mathematically from the central limit theorem, but
numerical experiments (section 3) support its use when many modes are
used.

The covariance matrix involves the eight-point correlation function,
computable from Wick's theorem and, for continuous and discrete fields 
by the methods of MVH97:
\ba
 & \langle\delta_1.....\delta_8\rangle_{\rm total}= & \nn
 & \langle\delta_1 \delta_2\rangle\langle\delta_3 
\delta_4\rangle\langle\delta_5 \delta_6\rangle\langle\delta_7 \delta_8\rangle + \ldots& 
\mbox{105 terms}   \nn
 & +\langle\delta_1 \delta_2\rangle\langle\delta_3
\delta_4 \delta_5 
\rangle\langle\delta_6\delta_7
\delta_8\rangle  +  \ldots &\mbox{280 terms} \nn
 & +\langle\delta_1 \delta_2\rangle\langle\delta_3
\delta_4 \delta_5 \delta_6\rangle\langle\delta_7
\delta_8\rangle  +  \ldots &\mbox{210 terms} \nn
 & +\langle\delta_1 \delta_2 \delta_3\delta_4\rangle\langle\delta_5 \delta_6 \delta_7\delta_8\rangle  +
\ldots & \mbox{35 terms}  \nn
 & +\langle\delta_1 \delta_2\rangle
\langle\delta_3\delta_4\delta_5 \delta_6 \delta_7\delta_8\rangle + \ldots & \mbox{28 terms} \nn
 & +\langle\delta_1 \delta_2\delta_3\rangle
\langle\delta_4\delta_5 \delta_6 \delta_7\delta_8\rangle + \ldots & \mbox{56 terms}   \nn
&  +\langle\delta_1\ldots \delta_8\rangle. &
\label{cov}
\ea
These expressions apply both to continuous fields and discrete point
processes, and the averages on the r.h.s. are all connected parts
only.  Under the hypothesis of GIC, as long as linear theory
holds\footnote{As shown above, second order contributions vanish for
gaussian primordial fluctuations, therefore linear theory breaks down
when third-order contributions become important.}, all
terms involving the connected $m$-point correlation with $m$ greater
or equal than two are non-zero only in the presence of shot noise.
Moreover for equal-side configurations the terms involving
$\langle\delta_l \delta_m \delta_n\rangle$ are also zero, since one
cannot make a triangle from any 3 of the wavevectors.  For clarity we
neglect shot noise in the main text, but the calculations have been
performed including it.

For GIC, the only non-zero terms come from the 
two-point function, i.e. the power spectrum.
We can test the gaussian hypothesis by including only these
gaussian terms in the covariance matrix, reducing the process to $\chi ^2$ analysis. This should also be a good
approximation for mildly non-gaussian fields (cf.  \cite{Hea98} 
for the microwave background).  With this covariance matrix,
a measurement of $\tau$ which is significantly non-zero would only rule
out gaussian statistics, but would not necessarily recover the correct 
value of $\tau$, which would require use of the correct {\it non-gaussian}
covariance matrix.  

\subsection{A priori estimation of the error on $\tau$}

We can readily calculate the expected variance of our estimator for
$\tau$.  The calculation follows MVH97 so we only sketch the calculation
here.  Ignoring shot noise, for square configurations we have
\be
\mu_{\alpha}=\langle D_{\alpha} \rangle = (2 \pi)^6 P^2 (\delta^D)^2+(2\pi)^3 \tau P^2\delta^D
\ee
and the variance is:
\be
\sigma_{\alpha}^2=\langle D_{\alpha}^2 \rangle = 4 P^4 \left[(2 \pi)^3 
\delta^D\right]^4
\ee
since only 4 terms are non-zero out of the 105.  Hence
\be
\sigma_{\tau}^{-2}=-\langle\frac{\partial^2 \ln{\cal L}}{\partial
\tau^2}\rangle_{\tau=0}=\sum_{\alpha}\frac{\mu_{\alpha}(\tau=0)P^2}{\sigma_{\alpha}^2}.
\ee 
The number of uncorrelated squares in a thin shell in $k$-space of
width $\delta(\ln k)$ is $\pi k^3 g \delta(\ln k)$ where
$g=V/(2\pi)^3$ is the density of states.  Considering contributions
from all the shells to the continuum limit we obtain
\be
\sigma_{\tau}^{-2}=\frac{1}{2^3\pi^2}\frac{1}{4V}\int_{k_{min}}^{k_{max}}k^3d(\ln k).
\label{scaling1}
\ee
$k_{max}$ is set by the breakdown of second-order perturbation theory
and $k_{min}$ by the size of the sample.  Therefore the error on
$\tau$ scales as $\sqrt{V}$ and the maximum signal-to-noise is
obtained by splitting the volume into subunits. This procedure to
reduce the error when dealing with higher-order spectral statistics is
widely used in other fields such as signal processing\footnote{Where it is
referred to as {\it segment averaging}.} and fluid
mechanics see e.g., \citet{Brillinger1}, \citet{Brillinger2},
\citet{Lii76} and has been already applied in cosmology by
\citet{MVH97} and \citet{VHMM98}.  This appears to give something for
nothing, but of course this is not true.  The resolution of the
paradox is given in section 2.3.

The minimum size of the samples may be set by the breakdown of
perturbation theory, or possibly by aliasing difficulties if there is
significant power on scales larger than the box.

Note that the analysis in
this section, and therefore the scaling (\ref{scaling1}), holds only as
long as the field is close to gaussian  so that non-gaussian elements
of the covariance matrix can be neglected.  As we said before, if
the field is close to gaussian we are justified in adopting a {\it
gaussian} likelihood, however if the distribution appears to be highly
non-gaussian this error calculation and volume dependence is no longer valid.
Since effectively $P(k)$ is estimated from the data themselves, in principle
one should propagate any error on $P(k)$ through into the error on
$\tau$. However, since the cosmic variance for the trispectrum is much larger
than that on the power spectrum, this effect should be
negligible. We confirmed this with numerical experiments.

The square configuration we considered above is not the only possible one. We
could also consider other configurations in which all the $k$-vectors
in the quadruplet have the same modulus. These configurations can be
parametrized by the angle $\theta$ between the first two vectors and
$90^\circ \ge \theta \ge 0^\circ$: in the square case
$\theta=90^\circ$, in the `degenerate' case $\theta =0^\circ$.  The above 
error analysis holds for all $\theta >0^\circ$.  For
the `degenerate' case of $\theta =0^\circ$, there are modifications to
the number of cyclic terms contributing to the means and covariance
matrix, and also to the number of independent quadrilaterals.  In this
case, we have $\mu_{\alpha}=2 P^2 [(2\pi)^3\delta^D]^2+(2\pi)^3\tau P^2\delta^D$,
$\sigma_{\alpha}^2=24 P^4 [(2\pi)^3\delta^D]^4$, and the number of uncorrelated
quadruplets is $2 \pi k^3 g \delta(\ln k)$.  These reduce the error by a
factor $\sqrt{2
\times 2^4/24}=1.15$, if $k_{max}$ remains the same.

\subsection{Volume-splitting  procedure}

Volume-splitting appears to improve the signal-to-noise.  We show
here in outline that it gives no more information than relaxing the fixed
quadrilateral shape and including more modes.

Consider the $n$ point spectrum and imagine that we wish to compute the
expected error on the amplitude $A$: $\langle \delta_{k_1}\ldots
\delta_{k_n}\rangle=A S\delta^D$ where $\delta^D\propto V$ and
$S$ is some fixed function. The leading term in the covariance,
neglecting shot noise, will be $\sim P^n(\delta^D)^n$. As already seen
before the density of states is $ g
\propto V$. If we fix the configuration of the $k$-vectors, then the
expected error on $A$, for a thin shell in
$k$-space, will be given by:
\be
\sigma^{-2}_{A}\propto \frac{g S^2(\delta^D)^2}{P^n(\delta^D)^n}\;
\ee 
which implies the curious result that $\sigma_{A}\propto
V^{(n-3)/2}$.  However, if we allow the shape of the $k$-vector
configuration to change we obtain:
\be
\sigma_{A}^{-2}\propto \frac{g^{(n-1)}S^2(\delta^D)^2}{P^n(\delta^D)^n}
\ee
where the exponent $n-1$ of the density of states comes from the presence of
the Dirac delta function, i.e. the fact that the
polygon has to close and therefore has $n-1$ degrees of freedom.
In this case we obtain the (more intuitive result) that:
$\sigma_{A}\propto 1/\sqrt{V}$ when all shapes are allowed for.

In the case of the power spectrum (as in the example of
\citet{NUMREC}) the two possibilities coincide ($n$=1), for this
reason splitting the volume does not improve the errors.  Note also
that in the bispectrum case ($n$=3) of \citet{MVH97} and
\citet{VHMM98}, the error is independent of the size of the sample
because the triangle shape is fixed.  For the trispectrum ($n$=4)
$\sigma_{A}\propto \sqrt{V}$.  Here, as in all cases, one has the
choice between considering all shapes, or keeping the shape fixed and
subdividing.

\section{N-body tests} %
In this section, we perform tests with real-space results from N-body
simulations.  We analyze square configurations, and ensure that no
wavevector occupies more than one square, so the covariance matrix is
approximately diagonal (cf. MVH97 for the bispectrum).  The simulation is a $128^3$ particle, 100 $h^{-1}$ Mpc side
box, CDM-like simulation from the Hydra consortium \citep{Hydra},
with parameters $\Omega=1.0$, $\Lambda=0.0$, $\sigma_8=0.64$,
$\Gamma=0.25$, and GIC.  Shot noise is
completely negligible in all scales of interest ($k < 1$; here and
hereafter $k$ is in units of h/Mpc).  

The breakdown of linear theory for the trispectrum is not known in
advance, but from MVH97 one can expect that the leading-order
corrections for the trispectrum (i.e. contributions from the third
order in perturbation theory) should be small at least up to
$k=0.55$ h Mpc$^{-1}$.  A likelihood computation (cf., section 2.1) shows that for this square configuration,
linear theory for the 4-point function should be valid up to 
$k_{\rm max}=0.67$, which sets the upper limit for our further
analysis, giving 1250 squares in total.  This limit is also apparent 
in Fig. 1, which plots \footnote{The notation here
follows MVJ97: $D_{\alpha}$ is a single measurement of a
statistical quantity i.e. the trispectrum for a given (e.g., square)
$k$ configuration on a given scale (e.g., $k$=0.5).  $\langle D_{\alpha}\rangle$ is the ensemble 
average of this quantity. Since we assume that the volume under consideration
is a ``fair sample'', then the average of all  $D_{\alpha}$ in the simulation on
the same scale and with the same $k$ configuration is an estimate of  $\langle
D_{\alpha}\rangle$.}$\langle D_{\alpha}\rangle$.
For this wavenumber limit, $\chi^2$ is shown as a function of $\tau$ in
Fig. 2: the minimum reduced $\chi^2$ is 0.8 and $\tau=(-1.5 \pm 5.5) \times
10^4 $ is consistent with zero.
The a priori estimation for the error of Eq.(\ref{scaling1}) yields
$\sigma_{\tau}=5.6 \times 10^4$.
Note that the estimate of $\tau$ is independent of linear biasing because
of the presence of the same number of $\delta$ factors in the numerator and
denominator of Eq. 10.

We have also analyzed another N-body simulation with very non-GIC.  The initial conditions were set by applying the following
mapping to a gaussian field:
\be
\delta \longrightarrow \delta^2 - \langle\delta^2 \rangle
\ee
(the so-called $\chi^2$ model for initial conditions).
The linear power spectrum is a power law with spectral index $n=-1$.
The simulation we used has $80^3$ particles, in a 200 $h^{-1}$ Mpc
side box, with $\sigma_8=0.5$. In this simulation the shot noise is
not negligible, but it can be included with the methods of MVH97.
This family of initial conditions could arise from evolution through
inflation of an isocurvature CDM model for structure formation and is motivated by the fact that it
accommodates galaxy formation at high redshift (e.g.,
\cite{Pee99a,Pee99b}).  Despite the low
normalization, in this case we find that for all $k$-space shells with
$k<0.67$ the minimum $\chi^2$ value for $\tau$ is nonzero at high
significance: considering all $k$-vectors in the range $0.15<k<0.67$
for equilateral configurations and in the range $0.15<k<0.75$ for
degenerate configurations, we obtain $\tau_{eq}=(9.2 \pm 1.6)\times
10^5$ and $\tau_{deg}=(6.4\pm 1.3)\times 10^5$ respectively.  Thus we
can {\em reject} the GIC hypothesis with confidence, but note that for this
highly non-gaussian model the covariance matrix is not accurate, so we
can not rely on the actual {\em measurement} of $\tau$ being reliable.

Incidentally, for this simulation, the bispectrum also differs in the
mildly non-linear regime from the one that would have originated by
gravitational instability from GIC. However, at the bispectrum level,
there is degeneracy between biasing and initial non gaussianity: in
practice it would not be possible to assess if there is substantial
bias/anti-bias or if the initial conditions were truly
non-gaussian. As we have mentioned, the trispectrum method is
bias-independent, at least as long as the bias is not strongly non-linear.

\subsection{Subdividing the Volume}

We have tested the subdivision procedure with static simulations of a
non-gaussian density field derived from a gaussian $\delta(x)$
\be
\delta(x)\longrightarrow \delta(x)+\epsilon \delta^2(x)
\label{map}
\ee
where $\epsilon$ is some parameter.  For simplicity, we take a white
noise power spectrum between $k_{\rm min}$ and $k_{\rm max}$.  Details of
the resulting 2-point and 4-point functions are given in the Appendix,
but they are summarized as
\be
P(k) = P_g(k)+2\epsilon^2 P_g(k)^2 V_k/(2\pi)^3
\ee
where $P_g(k)$ is the power spectrum of the underlying gaussian
$\delta$ field, and $V_k$ is a $k$-space volume defined in the
appendix.  The 4-point function is
\be
\langle
\delta_{\vk_1}\delta_{\vk_2}\delta_{\vk_3}\delta_{\vk_4}\rangle_{{\rm
conn.}}\sim(48\epsilon^2 P_g^3+\epsilon^4 P_g^4 V_k/(2\pi)^3)V.
\ee
We choose $P_g=2(2\pi)^3$ in the range $0.1<k<0.837$ and zero
elsewhere, and $\epsilon=0.4$.  In this case, the non-gaussian terms
in the covariance matrix are small in comparison with the gaussian one
and we ignore them.  We split the volumes into cubes of side $20$ and
$24\,h^{-1}$ Mpc, and make repeated simulations to reduce errors.  In
all, we analyzed 460 of the smaller boxes and 266 of the larger boxes,
so the total volume is the same in the two cases.  The results are
shown in Fig. {\ref{scaling2}}.  The expected value of $\tau$ is 1040,
and is recovered within the errors.  Note also that the errors (160
and 250 respectively) scale as expected $\propto N$ where $N$ is the
number of boxes, and the signal-to-noise is indeed higher for the
ensemble of smaller boxes, as expected.

\section{Complications with a real survey} %

In section 3 we showed that the trispectrum is a useful discriminant
between GIC and non-GIC in a very
idealized case where the field is unbiased, and the positions of the
particles are known in real space.
Note however that galaxy catalogues have an average number of galaxies
per unit volume that is about two orders of magnitude smaller than the
one of the simulation used, and use the redshift as a third spatial
coordinate. The resulting redshift-space map of the galaxy
distribution is therefore distorted and shot noise is significant. This is not a trivial issue since
we are pushing into the mildly non-linear regime, which is
significantly affected by non-perturbative contributions.

In this section, we show how to deal with these effects.

\subsection{Redshift-space distortions}

It is a convenient approximation to split redshift-space distortions
into two components \citep{Kaiser87}: a large scale distortion
responsible of the squashing known also as the `bull's eye' effect and
a small-scale radial smearing responsible for `Fingers-of-God'.  The
large-scale effect on individual Fourier components of the density
fluctuation is well described by multiplication by the Kaiser
factor $(1+\beta\mu^2)$, where $\mu$ is the cosine of the angle
between the $k$-vector and the line of sight, and $\beta\equiv
\Omega_0^{0.6}/b$ with $b$ the linear bias parameter.

The small-scale effect is hard to treat exactly and could potentially
erase the signal we are trying to detect; in fact virialized motions on small
scales produce a radial smearing and the associated Finger-of-God
effect contaminates the wavelengths we are interested in.  A
successful model that fits the power spectrum in numerical simulation
reasonably well (e.g., \cite{HC98,PD94}) assumes that the small-scale
velocity field is uncorrelated with density and has an exponential
velocity distribution.  Although not rigorously
theoretically-motivated, it has been shown in VHMM98 that this
modeling of the small-scale velocity dispersion works well as an
addition to perturbation theory.  In the Fourier domain, and in the
distant observer approximation, the exponential velocity dispersion
gives a damping factor, that, combined with the boosting factor of the
large-scale effect \citep{Kaiser87} gives \citep{PD94}:
\be
\delta_{\bf{k}} \longrightarrow
\delta_{\bf{k}}\frac{(1+\beta\mu^2)}
{\sqrt{1+k^2\sigma^2\mu^2/2}},
\label{veldisp}
\ee
where $\sigma$ is the pairwise velocity dispersion of galaxies.  With
this model for redshift-space distortions:
\be
\langle D_{\alpha}\rangle(k) \longrightarrow \langle D_{\alpha}
\rangle(k,\mu)=\langle D_{\alpha}\rangle(k)\frac{(1+\beta\mu^2)^4}
{\sqrt{\prod_{i=1,..,4}(1+k_i^2\sigma^2\mu_i^2/2)}}.
\label{tred}
\ee
As in VHMM98 we allow $\sigma$ to be scale-dependent (to fit the power
spectrum -- there is some observational evidence for this; see
\citet{HamTeg2000}) and we reject the $k$-vectors aligned too closely
with the line-of-sight (see \citet{SCF99} for an alternative model).

We have tested the model (\ref{tred}) for square configurations by
performing a $\chi^2$ analysis for the parameter $\tau$ on the
unbiased redshift-space catalogue created from the N-body simulation
with GIC.
 
Following VHMM98 the covariance matrix is modified as follows. The
power spectrum is first replaced by $P({\bf k}) \equiv P(k)(1+ \beta
\mu^2)^2$ and the resulting covariance matrix is then divided by
$\sqrt{\prod_{i=1}^{8}(1+k_i^2 \sigma^2 \mu_i^2/2)}$, where the index
$i$ runs over the 8 $k$-vectors that form the two squares.

The knowledge of the real-space power spectrum is required in this
model: in a realistic application (like for example Sloan and 2dF) an
accurate fit for the galaxy real-space power spectrum will be
known. Thus for the present work we shall assume that the real-space
power spectrum is known: only the velocity dispersion $\sigma$ needs
to be determined, and this can be done by fitting the power
spectrum. The limit of validity of the small-scale redshift-distortion
model for the trispectrum can also in principle be determined,
although in this case the limiting factor is the breakdown of
perturbation theory.  A likelihood analysis of the redshift-space
Fourier modes can give $\sigma$ in a similar manner to VHMM98.

The result for the redshift-space analysis is shown in
Fig. {\ref{figchired}}. Again the true value for $\tau$ is well
recovered within the errors: $\tau=(-1.0 \pm 5.5)\times 10^4$ (reduced
$\chi^2$ value is 0.95) for the square configuration and $\tau=(0.8\pm
4.5)\times 10^4$, for the degenerate configuration in good agreement with
the predicted error of $4.3\times 10^4$ obtained as outlined in
section 2.2 .

As a test of the velocity dispersion, we consider the
quadrupole-to-monopole ratio of $\langle D_{\alpha}\rangle$, following its use with the power
spectrum \citet{Ham92} and the bispectrum (VHMM98,
\citet{SCFFHM98}).  For the degenerate configuration all vectors have the same
$\mu^2$, so the quadrupole-to-monopole ratio can be simply defined by 
$R_T \equiv \langle D_{\alpha}\rangle^{(2)}/\langle D_{\alpha}\rangle^{(0)}$,
where the quadrupole and monopole moments of $T$ are
\ba
\langle D_{\alpha}\rangle^{(2)}& = & \frac{5}{2}\int_{-1}^{1} \langle D_{\alpha}\rangle(k,\mu)(3\mu^2-1)d\mu \nn
\langle D_{\alpha}\rangle^{(0)} & =&\frac{1}{2}\int_{-1}^{1} \langle D_{\alpha}\rangle(k,\mu)d\mu. 
\ea
In Fig. \ref{ratiodeg} we show how the model (\ref{tred}) reproduces
the observed $R_T$ in the GIC simulation.  We follow the same
procedure as VHMM98, where $\sigma$ is allowed to be slightly
scale-dependent, constrained to fit the power spectrum, and $k^2\mu^2
\le 0.3$.  It would be possible to define a similar quantity also for
non-degenerate configurations, but in this case $\langle D_{\alpha}\rangle$ in redshift space
would depend on two different angles to the line-of-sight which leads
to more complexity.

For the non-gaussian isocurvature field of Section 3.1, we find a $\tau$ which
is inconsistent with zero: e.g., for degenerate configurations  $\tau_{deg}=(6.4\pm 1.3)\times10^5$.  For non-gaussian models,
in general, the signal $\tau$ will be shape-dependent, (as for example
in the particular case illustrated in section 5). This
dependence on the $k$-vector configuration might therefore hold some extra information or
allow stronger detection of a non-gaussian signal.

\subsection{Expected performance from PSCz and Sloan}

The PSCz survey \citep{SPSCZ98} is the largest nearly-all-sky survey,
containing around 15000 galaxies.  The space density of galaxies is
not very high, so shot noise is important beyond about 40 $h^{-1}$ Mpc 
for $k=0.67$.  Assuming the initial field is close to gaussian, we use 
the gaussian covariance matrix, but with shot noise included.
  
Before being able to perform the trispectrum analysis on this survey
however still there are a number of unresolved issues. The most
important are the practical effects of the rapidly-varying selection
function and the radial nature of the redshift-space distortions. In fact the
technique described in section 4.1 assumes that the spherical nature
of the distortion can be neglected and the sky can be considered flat
(the so-called `distant-observer approximation', or `plane-parallel
approximation'). We must emphasize that this approximation does not
hold for the PSCz, and one really has to go beyond the plane-parallel
approximation, by using spherical harmonics for example
(e.g., \citet{HT95}).  

In the case of the 2dF and Sloan surveys, due to the bigger volume
available, the problem of the radial nature of the redshift-space
distortions is much reduced, and the selection function issue can be
tackled as illustrated in the appendix. The 1$\sigma$ error achievable on $\tau$ is very
encouraging: $1 \times 10^{4}$ for Sloan, if we restrict the analysis
to the linear regime ($k<0.3 h$) and divide the survey conservatively
into 100 $h^{-1}$ Mpc side cubes.  This reduces to $4 \times 10^{3}$
if wavenumbers up to $k=0.7 h$ Mpc$^{-1}$ are included. The error can be
reduced even further by considering smaller boxes.

\subsubsection{Meaning of non zero value  for $\tau$ } 

Most inflationary models predict deviations from gaussianity of the form
described by equation (\ref{map1})--applied to the potential rather than the
density fluctuation field; this kind of non-gaussianity affects
mainly the bispectrum (e.g., \cite{VWHK00}). An order of magnitude calculation
shows that for the non-gaussian models
analyzed in \cite{VWHK00}, the minimum deviation from gaussianity that can be
detected with the large-scale-structure bispectrum is at least a factor of a few smaller than the one
detectable with the large-scale-structure trispectrum; in these cases an analysis of the
cosmic microwave background bispectrum is thus more promising. However, as already mentioned, primordial
non-gaussianity might arise on physical scales which
are difficult to probe with the microwave background, the large-scale-structure  bispectrum
might deviate from the perturbation theory prediction due to the presence of
bias or antibias, and non-gaussian models might have no primordial
bispectrum, or even negative initial bispectrum. It is nevertheless clear that
other plausible non-gaussian models such as the $\chi^2$ model described in
section 3 can easily be ruled out with a trispectrum analysis of Sloan/2dF data sets.

\section{Conclusions}

We have presented a method for using the trispectrum, the four-point
function in Fourier space, to discriminate between GIC and non-GIC
from linear large-scale structure data.  The advantage of this method
is that for mildly non-gaussian fields the analysis in real space is
independent of the power spectrum normalization, linear biasing, and
cosmology, and, as a spectral method,
the covariance matrix is more simply computed, in contrast to $n-$point
correlation functions.  In
redshift space, cosmology and bias enter only through the measurable
quantity $\beta$, we show how to deal with redshift-space
distortions (in the distant-observer approximation), we include in the
calculations effects of shot noise following the method presented in MVH97
(although for clarity the shot noise contribution is ignored in the text and
reported in the appendix) and
it is straightforward to model the effects of  varying selection
function using the method of MVH97 (see the Appendix).  The non-linear evolution of the trispectrum
is expected to be rather weak in perturbation theory, so our
expectation is that linear theory should provide a good description up
to scales where the field becomes significantly non-linear.  This is
born out (and quantified) in simulations. Fig. (1) illustrates this
point. The equilateral configurations bispectrum for the same
simulation considered here, agrees with second-order perturbation
theory up to $k\sim 0.55$ (MVH97). Fig (1) shows linear perturbation
theory is adequate for square trispectrum configurations up to $k \sim
0.67$.  We parameterize the departures from gaussian statistics by
introducing the related quantity $\tau$, which is the 4-point
correlation function in units of
$\sqrt{P(k_1)P(k_2)P(k_3)P(k_4)}$. This quantity, in specific cases,
can give us a meaningful measure of `non-gaussianity'.  We compute the
gaussian variance of $\tau$, which allows a test of the gaussian
hypothesis.  The error on $\tau$ can be computed straightforwardly,
and is found to be in good agreement with internal error from N-body
simulations.  For mildly non-gaussian fields, we can expect that the
use of a gaussian covariance matrix is still quite adequate, and in
such cases a {\em measurement} of $\tau$ can reliably be made, subject
to the requirement that the bias is linear on large scales.  For
highly non-gaussian fields, the gaussian hypothesis can be rejected,
but the measurement of $\tau$ will be unreliable as the covariance
matrix we use may neglect important terms.

Application to data will probably have to wait for completion of the
2dF and Sloan surveys, as the distant-observer approximation is not a
good one for the nearly all-sky PSCz survey.  In addition, the high
shot noise in the PSCz survey means that the expected errors on $\tau$ are
large enough to make it hard to distinguish between GIC and some
significantly non-gaussian fields.  The 2dF and Sloan surveys should
do much better. One could compute $\tau$ from volumes of side 100
$h^{-1}$ Mpc with a random error of less than one per cent for a
$\chi^2$ model as the one considered here, and could therefore put
tight constraints on non-gaussian models.  It would be helpful to
apply this method to defect models, but unfortunately none is
available at the required stage of evolution.

The major uncertainty in this analysis is the effect of bias.  A
non-linear bias term acts like a non-gaussian initial
field, and, as we are using linear perturbation theory here, we cannot
use polygons of different shape to lift the degeneracy, as done by
MVH97 to lift the degeneracy between bias and gravitational evolution
in the bispectrum.  The best hope is to constrain a combination of
initial non-gaussianity and quadratic bias, and to argue that if
the measured value of $\tau$ is zero, then it would require a conspiracy
unless both effects were absent.  In principle it is possible to
distinguish these two effects by considering differences at higher
order in perturbation theory, as an Eulerian non-linear bias is applied
to the evolved field, whereas a primordial non-gaussian field of
the same mathematical form is applied to the initial field, and thus
the evolution is different.  It is an open question as to whether any
realistic survey would have the signal-to-noise to do this.

\section*{APPENDIX}

We consider the transformation of a gaussian field as follows:
\be
\delta(x)\longrightarrow \delta(x)+\epsilon \delta^2(x)
\label{map1}
\ee
This can be seen as the first two terms of a Taylor expansion of any
non-gaussian field originating as a local mapping from a underlying gaussian 
one.

We choose a toy model for the power spectrum of the underlying
gaussian field $P_g$: a top hat function between some $k$ minimum and
maximum.  The power spectrum for the resulting field will then be:
\ba
P(k)=P_g(k)+\frac{2}{(2\pi)^3}\epsilon^2\int P_g(k^{\prime})P_g(\mid
\vk-\vk^{\prime}\mid)d^3k^{\prime}\sim\nn
 P_g(k)+2\epsilon^2 P_g(k)^2 V_k/(2\pi)^3
\ea
where $V_k$ is the volume where the integrand in the previous equation
is non-zero.  The relevant quantities for our analysis, substituting
the Dirac delta function by $V/(2 \pi)^3$ are the following:
\be
\langle
\delta_{\vk_1}\delta_{\vk_2}\delta_{\vk_3}\delta_{\vk_4}\rangle_{{\rm
conn.}}\sim(48\epsilon^2 P_g^3+\epsilon^4 P_g^4 V_k/(2\pi)^3)V,
\ee
\ba
{\rm SIGNAL}\equiv \tau\equiv \frac{\langle
\delta_{\vk_1}\delta_{\vk_2}\delta_{\vk_3}\delta_{\vk_4}\rangle_{{\rm
conn.}}V}{\langle\delta_{\vk_1}\delta_{\vk_2}\rangle\langle\delta_{\vk_3}\delta_{\vk_4}\rangle}\sim\nn
\frac{48
\epsilon^2 P_g^3+\epsilon^4P_g^4V_k/(2\pi)^3}{(P_g^2+4\epsilon^2 P_g^3V_k/(2\pi)^3+4 
\epsilon^4 P_g^4 [V_k/(2\pi^3)]^2)}
\ea

\ba
{\rm NOISE}\simeq \sqrt{(2
\pi)^3V}\frac{2}{\sqrt{k_{max}^3-k_{min}^3}}
\ea

The covariance matrix is dominated by the gaussian terms if $\langle
\delta_{\vk_1}\delta_{\vk_2}\delta_{\vk_3}\delta_{\vk_4}\rangle_{{\rm
conn.}} \ll \langle
\delta_{\vk_1}\delta_{\vk_2}\rangle \langle
\delta_{\vk_3}\delta_{\vk_4}\rangle$, but at the same time we want to maximize
the signal-to-noise ratio.

This considerations suggest that the best choice is: $P(k)=2 (2\pi)^3$ if
$0.1<k<0.837$ and zero elsewhere and $\epsilon=0.4$.  This allows us to
use the gaussian covariance matrix, but also to have a reasonable
signal-to-noise for the non-gaussian component from a single box of
size of a few tens of Mpc.

\subsection*{Shot Noise}
In the application of the method to a real survey the shot noise might
be important. Here we report how to modify the relevant expressions in
the presence of shot noise. In what follows $\overline{n}$ denotes the
average number of particles per unit volume, and the superscript $d$
indicates that these results apply to discrete, point processes.  We
follow the generating functional approach of MVH97 to find the
following:
\be
\langle\delta_i \delta_j\rangle^d_c \longrightarrow (2\pi)^3\left[ P(k_i)+
\frac{1}{\bar{n}} \right] \delta ^D({\bf k}_i + {\bf k}_j) 
\ee
\ba 
\langle\delta_l \delta_m \delta_n\rangle_c^d & \!\!\!\!\!\!\!\!\!\!\!\!\!\!\!\!\!\!\!\!\!\!\!\!\!
\longrightarrow(2\pi)^3 \delta^D({\bf k}_l+{\bf k}_m+{\bf k}_n)\times & \nn
& \!\!\!\!\! \left\{ B_{lmn} + \frac{1}{\bar{n}} \left[ P(k_l)+P(k_m)+
P(k_n) \right]+ \right.
\left. \frac{1}{\bar{n}^2 } \right\} & ,
\ea
where $B_{lmn}$ denotes the Bispectrum,
\ba
\langle\delta_o\delta_p\delta_q\delta_r\rangle_c^d  &\longrightarrow  & (2\pi)^3 
\delta^D({\bf k}_o+{\bf k}_p+{\bf k}_q+{\bf k}_r)\times \nn
& &\left\{ \frac{1}{\bar{n}} \left[ B_{(o+p) q r} + \mbox{perm. 
(6 terms)} \right] + \right. \nn
& &\frac{1}{\bar{n}^2} \left[ (P_{o+p+q} + \mbox{cyc. (4 terms)})+
\right.\nn
& & \left. P_{o+p}+P_{o+q}+P_{o+r} \right] +\nn
& &\left. \frac{1}{\bar{n}^3 } \right\} ,
\ea
\ba
\langle\delta_1\ldots\delta_5\rangle^d_c& \longrightarrow &(2\pi)^3 
\delta^D({\bf k}_1+{\bf k}_2+{\bf k}_3+{\bf k}_4+{\bf k}_5 )\times \nn
& &\left\{ \frac{1}{\bar{n}^2} \left[ B_{(1+2) (3+4) 5} + \mbox{perm. 
(15 terms)} \right] + \right. \nn
& &\left. \frac{1}{\bar{n}^2} \left[ B_{(1+2+3) 4 5} + \mbox{perm. 
(10 terms)} \right] \right\} + \nn
& &\frac{1}{\bar{n}^3} \left[ (P_{1+2} + \mbox{cyc. (10 terms)})+
\right.\nn
& & \left. +P_{1}+P_{2}+P_{3}+P_{4}+P_{5} \right] +\nn
& &\left. \frac{1}{\bar{n}^4 } \right\} ,
\ea

\ba
\langle\delta_1\ldots\delta_6\rangle^d_c& \longrightarrow & 
(2\pi)^3 \delta^D({\bf k}_1+ \ldots
+ {\bf k}_6) \times \nn
& &\left\{ \frac{1}{\bar{n}^3} \right.  [ B_{12(3+\ldots+6)}+ 
\mbox{perm. (15 terms)}+   \nn
 & &               B_{(1+2)(3+4)(5+6)}+\mbox{ perm. (15 terms.)}+ \nn
 & &               B_{1(2+3)(4+5+6)} +\mbox{perm. (60 terms)} ] + \nn
 & &  \frac{1}{\bar{n}^4 } [ P_1+\ldots+P_6 +\nn
 & &               P_{1+2} + \mbox{perm. (15 terms)}+ \nn
 & &               P_{1+2+3} + \mbox{ perm. (10 terms)}]+  \nn
 & & \left. \frac{1}{\bar{n}^5 } \right\}. 
\ea
\ba
\langle\delta_1\ldots\delta_8\rangle^d_c & \longrightarrow &  
(2\pi)^3 \delta^D({\bf k}_1+ \ldots + {\bf k}_8) \times \nn
 & &\left\{ \frac{1}{\bar{n}^5} \right. [B_{1 2 (3+\ldots+8)}+\mbox{perm. (28
terms)}+ \nn
 & &         \!\!\!  B_{1 (2+3) (4+\ldots+8)}+\mbox{perm.(168 terms)}+ \nn
 & &         \!\!\!  B_{1 (2+3+4) (5+\ldots+8)}+\mbox{perm.(280 terms)}+ \nn
 & &  \!\!\!\!\!\!\!\! B_{(1+2) (3+4+5) (6+7+8)}+\mbox{perm.(280 terms)}+ \nn
 & &  \!\!\!\!\!\!\!\!\! B_{(1+2) (3+4+5+6) (7+8)}+\mbox{perm.(210 terms)}]+ \nn
 & &  \frac{1}{\bar{n}^6 } [ P_1+\ldots+P_8 +\nn
 & &               P_{1+2} + \mbox{perm. (28 terms)}+ \nn
 & &               P_{1+2+3} + \mbox{ perm. (56 terms)}+  \nn
 & &               P_{1+2+3+4} + \mbox{ perm. (35 terms)}+  \nn
 & & \left. \frac{1}{\bar{n}^7 } \right\}. 
\ea
Neglecting shot noise when it is not negligible has two effects. The main
effect is to overestimate $H$.  E.g., the generalization of (\ref{Dalpha}) is
$T\rightarrow T_{SN}$, where
\ba
(2\pi)^3 T_{SN} = & \nn
(2\pi)^3 (P(k_1) & \!\!\!\!\!\! +
\frac{1}{\bar{n}})(P(k_3)+\frac{1}{\bar{n}})\delta^D({\bf k}_1+{\bf k}_2)
\delta^D({\bf k}_3+{\bf k}_4)  \nn
 +\langle\delta_1\ldots\delta_4\rangle^d_c &
\ea
The second effect is that the errors are underestimated.  In
particular, even in the gaussian case, there are additional connected
terms in the correlations arising from shot noise.

\subsection*{Selection function}
The above analysis is valid for volume-limited samples, where the mean
number density of galaxies is independent of position.  In a realistic
catalogue the presence of a varying selection function makes the mean
density position-dependent and might even induce spurious detection of
non-gaussianity. However it is possible to treat this effect
accurately by the generating functional approach as in
MVH97.  If $n(\bf{x})$ is the mean density of such a catalogue,
following
\cite{FKP94}, we can define a working fluctuation field composed by 
subtracting from the real catalogue a synthetic catalogue with no
clustering, but with the same selection function, and then weighting
suitably the combination:
\be
F({\bf x})=\gamma w({\bf x})[n({\bf x})-\alpha n_s({\bf x})] 
\ee
where $w({\bf x})$ is an arbitrary weighting function, $\gamma$ is a
normalization factor, $n_s({\bf x})=n({\bf x})/\alpha$, and $\alpha$
is the ``dilution'' of the synthetic catalogue. We will then consider
the limit $\alpha \longrightarrow 0$ to avoid shot noise in the
synthetic catalogue.  The $n$-point correlation functions in Fourier
space are easily calculated by considering $F$ as the superposition of
the process $f=\gamma w({\bf x})n({\bf x})$ and $f_s=-\alpha \gamma w({\bf
x}) n_s({\bf x})$.  The generating functional for $F$ will then be:
${\cal Z }_F({\cal J})={\cal Z}_f({\cal J}){\cal Z}_{f_s}(-\alpha{\cal
J})$.  Ignoring shot noise,
\be
{\cal Z}_{f_s}(-\alpha{\cal J})=- \alpha i \int d^3x
{\cal J}_s 
\ee
where ${\cal J}_s=\gamma w({\bf x})n_s({\bf x})\sum_{m}s_m \exp(- i
{\bf k}_m \cdot {\bf x})$, and the {\it ansatz\/ } for the generating
functional for the field $f$ is:
\begin{equation}
\begin{array}{c}
{\cal Z}_f[{\cal J}] =\\
\exp \left[ i \int d^3x {\cal J}(\vec{x})-\frac{1}{2}\int d^3x d^3 x^{\prime} 
{\cal J}(\vec{x}){\cal J}(\vec{x}^{\prime} ) \xi^{(2)}_{\rm conn.} 
(\vec{x},\vec{x}^{\prime} ) 
\right.\\
\left. - \frac{i}{6} \int d^3x d^3x^{\prime} d^3 x^{\prime \prime} {\cal J}(\vec{x}) 
{\cal J}(\vec{x}^{\prime} ) {\cal J}(\vec{x}^{\prime \prime} ) 
\xi ^{(3)}_{\rm conn.} (\vec{x},
\vec{x}^{\prime} ,\vec{x}^{\prime \prime} )\right.\\
\left. +\frac{1}{24}\!\!\int\!d^3\!x d^3\!x^{\prime} d^3\!\!x^{\prime \prime} d^3\!x^{\wr} {\cal J}(\vec{x}) 
{\cal J}(\vec{x}^{\prime}) {\cal J}(\vec{x}^{\prime \prime}\! )
\cdot\cdot\cdot{\cal J}(\vec{x}^{\wr} ) 
\xi ^{(4)}_{\rm conn.} (\vec{x},\cdot\cdot\cdot,\vec{x}^{\wr} ) \right]
\end{array}
\end{equation} 
Following the same procedure as in \cite{MVH97} it is possible to obtain 
the $n$-point function in Fourier space by differentiating the generating
functional; shot noise can be easily included by modifying the generating
functional .

The  $\tau$ in presence of a spatially-varying selection
function  therefore becomes: 
\be
\tau \longrightarrow \tau \frac{I_{44}}{I^2_{22}} 
\ee
where
\be
I_{ij}  \equiv  \int d^3{\bf x} w^i({\bf x}) n^j({\bf x})
\ee

In general the effect of a spatially-varying selection function can be
summarized as follows:

\ba
\gamma^2 & = & 1/I_{22}\\
V & \longrightarrow&  \frac{I_{NN}}{I_{22}^{N/2}}\;\;\;\; ; \;\;\;
\delta^D \longrightarrow \frac{I_{NN}}{(2 \pi)^3 I_{22}^{N/2}}\\
\frac{1}{\overline{n}^q} & \longrightarrow & \frac{I_{N(N-q)}}{I_{NN}}
\ea

\section*{Acknowledgments}
LV acknowledges NASA grant NAG5-7154 and TMR fellowship.
We are
grateful to Sabino Matarrese, Avery Meiksin, M. Strauss and the anonymous referee for helpful comments.
The simulation with gaussian initial
conditions was obtained from the data bank of cosmological N-body
simulations provided by the Hydra consortium
(http://coho.astro.uwo.ca/pub/data.html) and produced using the Hydra
N-body code \citep{Hydra}. LV thanks Alison Stirling for providing
the N-body simulation with non-gaussian initial conditions.  
 

\clearpage

{\bf Fig. 1.--}The 4-point correlation function in Fourier space
(disconnected and connected) $\langle D_{\alpha}\rangle\equiv\langle\delta_{\vk_1}\ldots\delta_{\vk_4}\rangle$ is approximated by averaging
square configurations with binned wavevectors from a CDM-like N-body simulation (solid
line). Also shown is the linear perturbation theory prediction for
gaussian initial conditions i.e.
 $\mu_\alpha(\tau=0)$ from equation (9), for a
square configuration of wavevectors.  Errors are errors in
the mean for each bin.  For this square configuration, linear perturbation
theory breaks down around $k=0.67 h$ Mpc$^{-1}$.

{\bf Fig. 2.--}Minimum $\chi ^2$ analysis for the parameter $\tau$ for the
gaussian CDM-like N-body simulation.  Only the `square' configuration
for the trispectrum has been considered here and $0.15 \le k \le 0.67$ (in
units of h  Mpc$^{-1}$). $\tau$ is a measure of the
connected part of the trispectrum; for gaussian initial conditions $\tau=0$
(vertical line). To produce this graph, we use
the gaussian initial conditions covariance matrix to compute the
likelihood (equation 11). See sections 2.1 and 3
for further details.

{\bf Fig. 3.--}The $\chi^2$ analysis for 460 volumes of $20 h^{-1}$ Mpc side
(dot-dashed line) and 266 volumes of $24 h^{-1}$ Mpc side (dashed
line) for $0.1< k <0.837$ (in
units of h  Mpc$^{-1}$). The $y$ axis is normalized so that 1-$\sigma$ limits are
at $y=1$.  The predicted value for $\tau$ is $\tau\simeq
1040$ (vertical lines).  This illustrates the
scaling with the volume of the  error on $\tau$ (see
Section 3.1 for more details).

{\bf Fig. 4.--}Minimum $\chi ^2$ analysis for the parameter $\tau$ from the 
redshift-space unbiased CDM-like N-body simulation (see text for
details). Only the square configuration  has been
considered here.  The value for the velocity dispersion parameter
$\sigma \simeq 980$km/s although it is slightly
scale-dependent.  The range of $k$-vectors considered is $k\le
0.67 h$ Mpc$^{-1}$.

{\bf Fig. 5.--} Quadrupole-to-monopole ratio for the trispectrum $R_T$ for
degenerate square configuration of a GIC simulation. The continuous line is
the theoretical $R_T$ using the redshift-space distortions as in VHMM98 (see
section 4.1 for details).

\clearpage

\begin{figure}
\begin{center}
\setlength{\unitlength}{1mm}
\begin{picture}(90,70)
\includegraphics{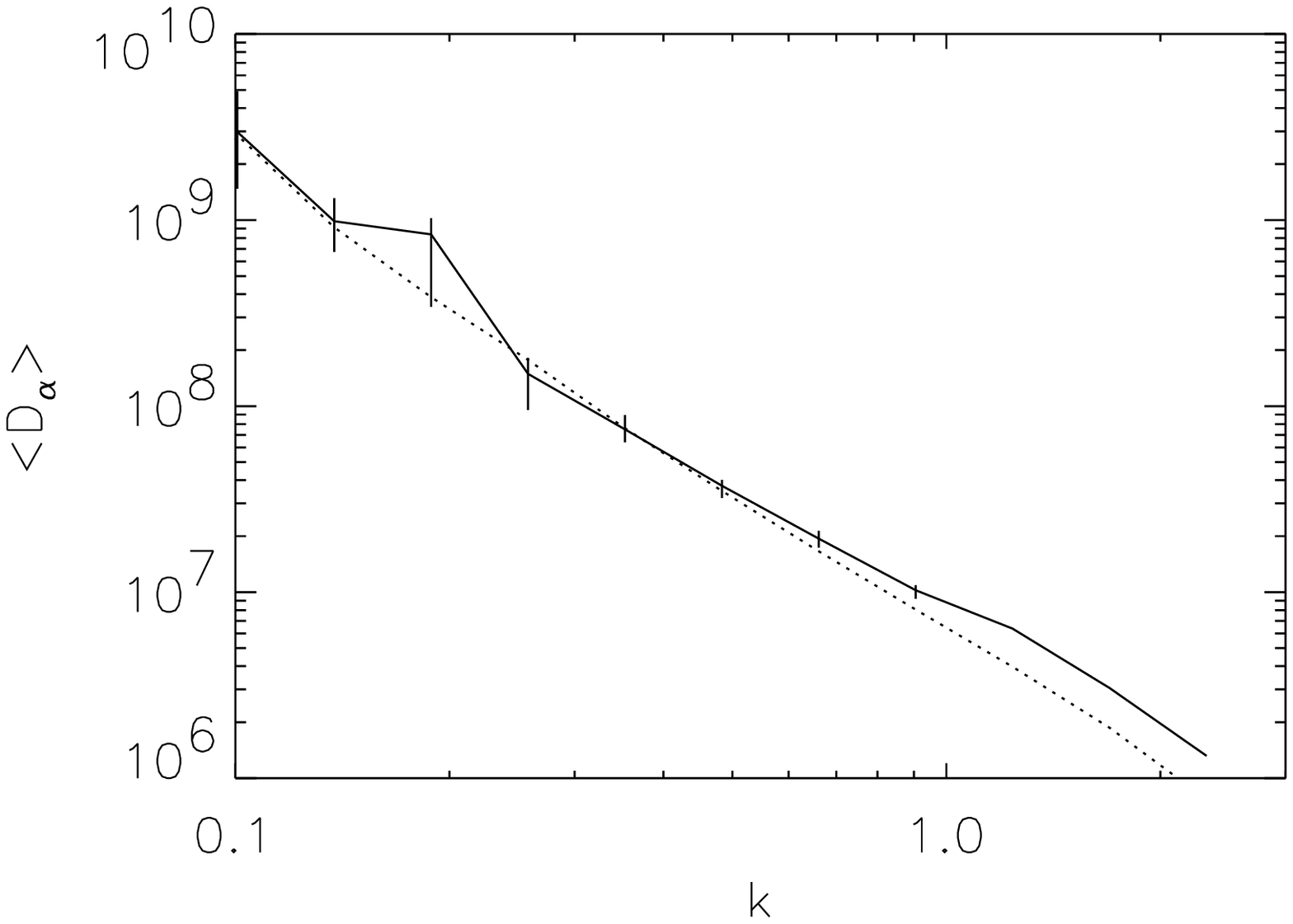}
\end{picture}
\end{center}
\caption{}
\label{figtrisp}
\end{figure}

\clearpage
\begin{figure}
\begin{center}
\setlength{\unitlength}{1mm}
\begin{picture}(90,70)
\includegraphics{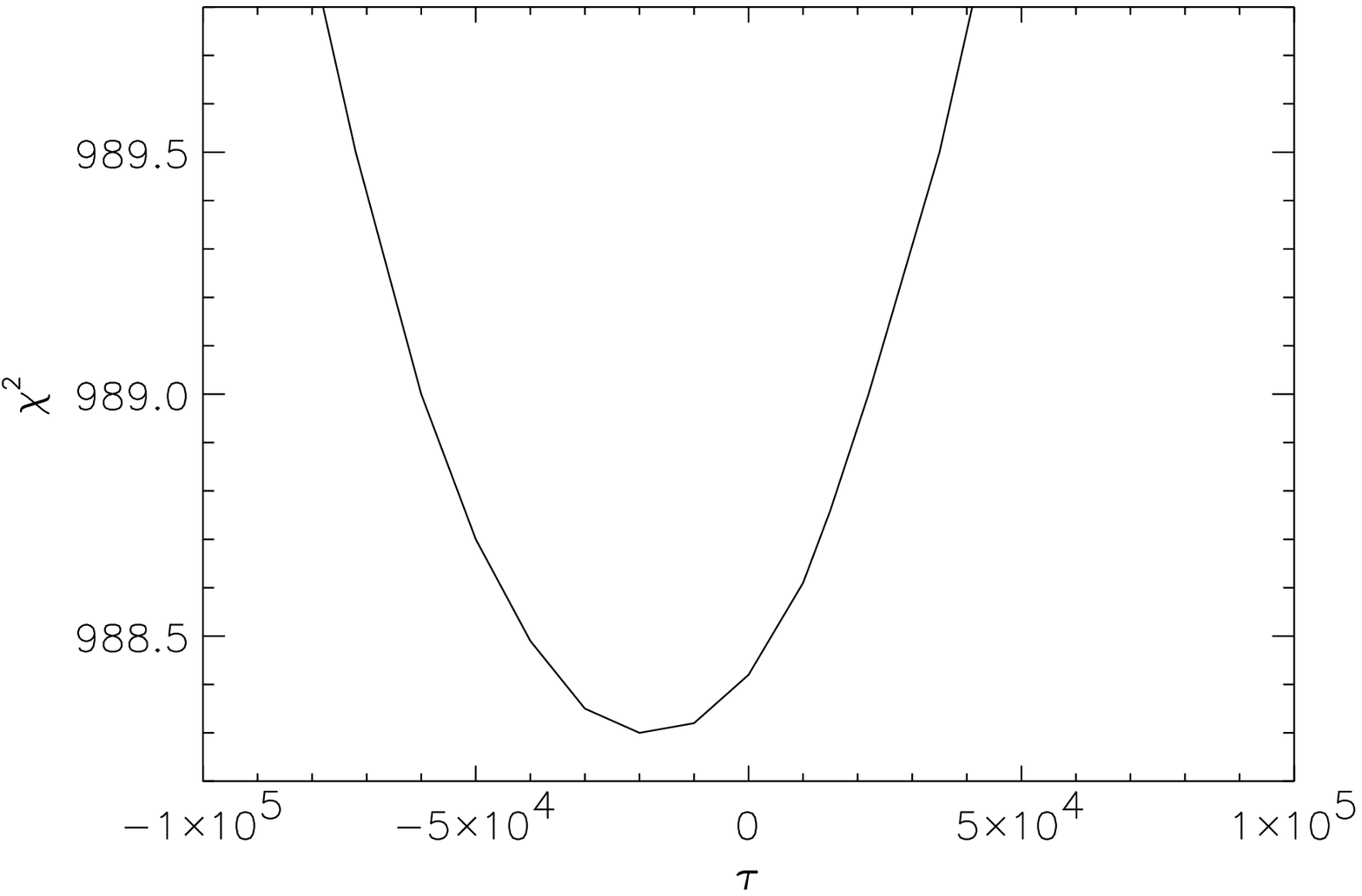}
\end{picture}
\end{center}
\caption{}
\label{figchireal}
\end{figure}

\clearpage
\begin{figure}
\begin{center}
\setlength{\unitlength}{1mm}
\begin{picture}(90,70)
\includegraphics{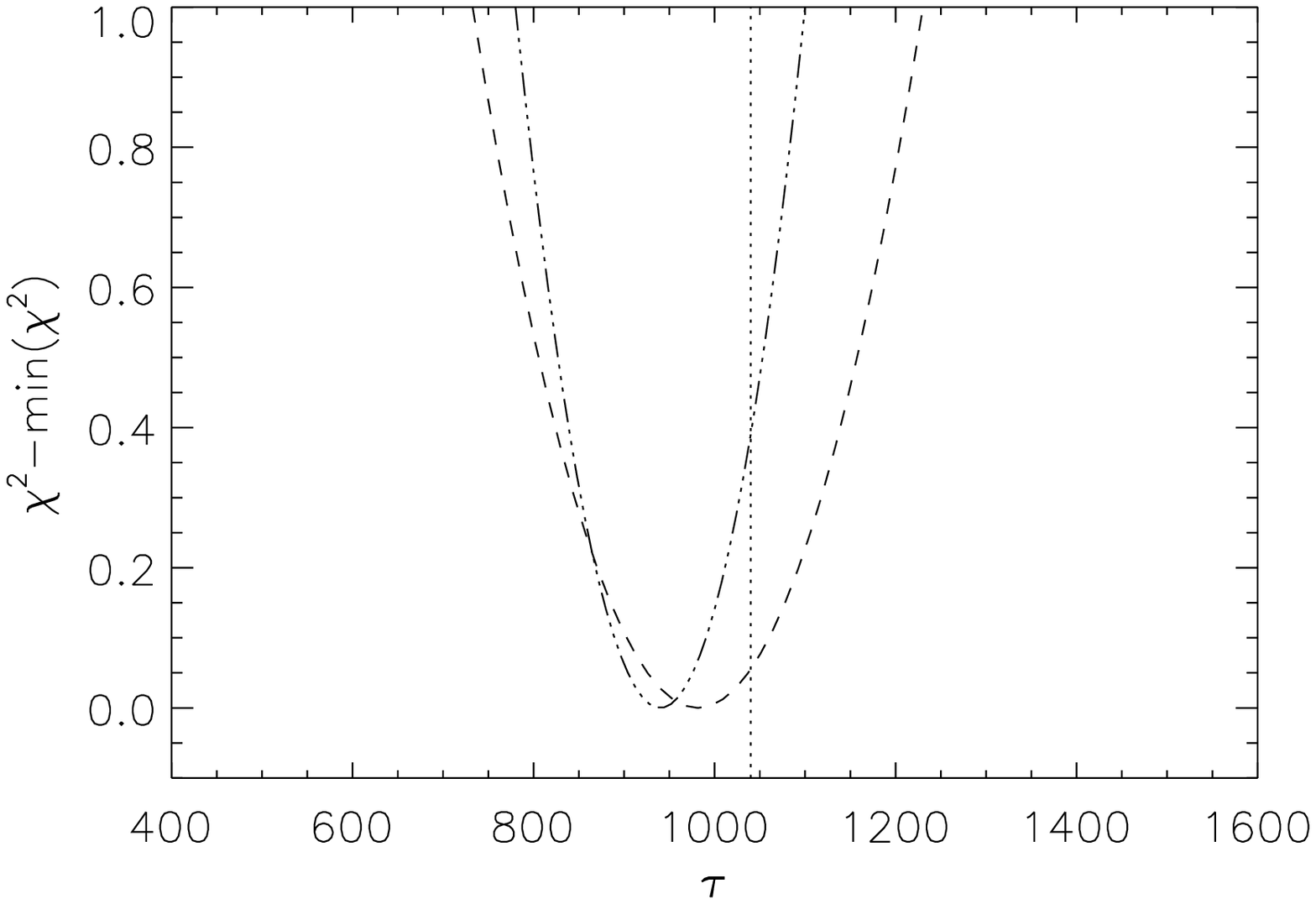}
\end{picture}
\end{center}
\caption{}
\label{scaling2}
\end{figure}

\clearpage

\begin{figure}
\begin{center}
\setlength{\unitlength}{1mm}
\begin{picture}(90,70)
\includegraphics{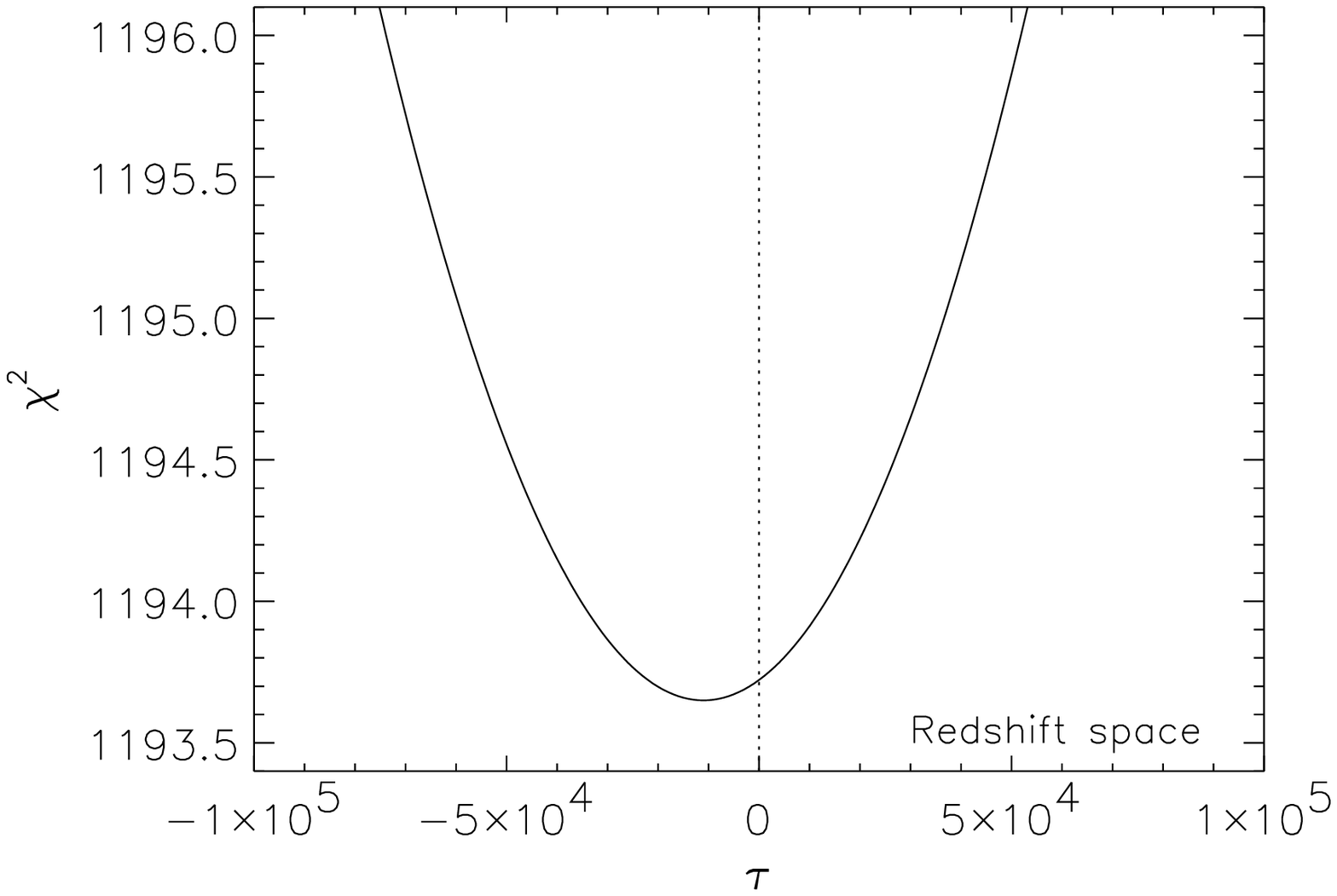}
\end{picture}
\end{center}
\caption{}
\label{figchired}
\end{figure}

\clearpage

\begin{figure}
\begin{center}
\setlength{\unitlength}{1mm}
\begin{picture}(85,60)
\includegraphics{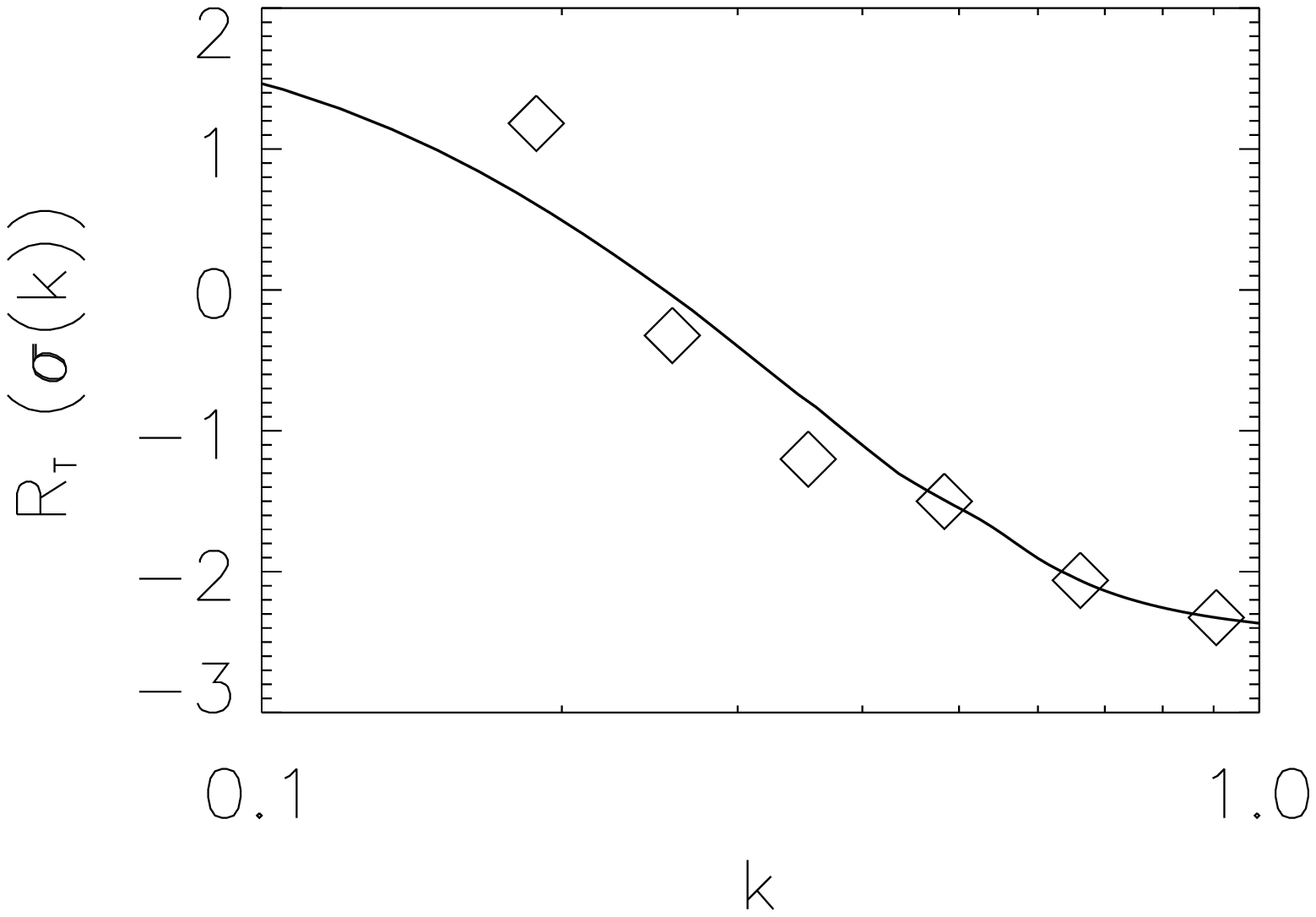}
\end{picture}
\end{center}
\caption{}
\label{ratiodeg}
\end{figure}

\end{document}